\documentclass{article}
\usepackage{spconf}
\usepackage{amsmath,graphicx,amsfonts, siunitx}
\usepackage[acronym]{glossaries}
\usepackage{xcolor}
\usepackage{pgfplots}
\usepackage{tikz}
\usepackage{graphicx}

\newacronym{CDF}{CDF}{cumulative distribution function}
\newacronym{PDF}{PDF}{probability density function}
\newacronym{WOSA}{WOSA}{Welch's overlapped segment averaging}
\newacronym[longplural={power spectral densities}]{PSD}{PSD}{power spectral density}
\newacronym{WP}{WP}{Welch percentile}
\newacronym{FFT}{FFT}{fast Fourier transform}
\newacronym{EDOF}{EDOF}{equivalent degree of freedom}

\newcommand\mycopyrighttext{%
  \footnotesize © 2021 IEEE. Personal use of this material is permitted. Permission from IEEE must be obtained for all other uses, in any current or future media, including reprinting/republishing this material for advertising or promotional purposes, creating new collective works, for resale or redistribution to servers or lists, or reuse of any copyrighted component of this work in other works.}
\newcommand\mycopyrightnotice{%
\begin{tikzpicture}[remember picture,overlay]
\node[anchor=south,yshift=10pt] at (current page.south) {\fbox{\parbox{\dimexpr\textwidth-\fboxsep-\fboxrule\relax}{\mycopyrighttext}}};
\end{tikzpicture}%
}

\title{Statistical Properties of a Modified Welch Method that uses Sample Percentiles}
\name{Felix Schwock, Shima Abadi\thanks{This research was supported by the Office of Naval Research grant number N00014-19-1-2644.}}
\address{Department of Electrical and Computer Engineering \\
    University of Washington, Seattle, WA 98195, USA}
\begin{document}
%

\maketitle

\mycopyrightnotice

\begin{abstract}
We present and analyze an alternative, more robust approach to the \gls{WOSA} spectral estimator. 
Our method computes sample percentiles instead of averaging over multiple periodograms to estimate \glspl{PSD}.
Bias and variance of the proposed estimator are derived for varying sample sizes and arbitrary percentiles. 
We have found excellent agreement between our expressions and data sampled from a white Gaussian noise process. 
\end{abstract}

\glsresetall

\section{Introduction}
\label{sec:intro}

The \gls{WOSA} method, first introduced by Welch in 1967 \cite{wel67}, is a popular approach for estimating \glspl{PSD} of stochastic signals due to its computational efficiency, its ability to scale estimation variance, and its potential to reduce spectral leakage.
However, the method can suffer from strong outliers in the data caused by transients or other broadband interfering signals.
Those outliers can prohibit an accurate estimation of the prevailing noise level, thus, limiting the scope of the \gls{WOSA} estimator \cite{mar19}.
A possible solution, which has proven to be successful in several spectral estimation applications (see for example \cite{all12, par09, mer12, kee17}), is to take the median of the periodograms at each frequency bin instead of the arithmetic mean.
This can be regarded as a special case of a more general Welch estimator that uses sample percentiles, in the following referred to as \gls{WP} estimator.
While the statistical properties of the classical \gls{WOSA} estimator have been analyzed thoroughly \cite{wel67, nut71, per20}, respective results for the percentile estimator are yet to be determined.
In this paper, we fill this gap by deriving formulas for bias, variance, and limiting distribution of the \gls{WP} estimator.
Section~\ref{sec:wosa_background} briefly reviews the concept of the classical \gls{WOSA} estimator and defines the \gls{WP} estimator. In
Section~\ref{sec:statistical_properties}, we derive the statistical properties of the \gls{WP} estimator.
Section~\ref{sec:simulations} compares the theoretical results with Monte Carlo simulations using data sampled from a white Gaussian noise process.

\section{Welch Percentile Estimator}\label{sec:wosa_background}

To compute \gls{PSD} estimates using Welch's method, the time domain signal sampled from a stationary process with sampling frequency $f_s$ is first divided into $K$ potentially overlapping segments, each of which containing $N_\mathrm{s}$ samples.
Each segment is then multiplied with a window function and the magnitude squared of their fast Fourier transform is computed.
The result is a set of modified periodograms $\lbrace \hat{P}_i(f_j)\rbrace_{i=1}^K$.
Therein, $f_j$ refers to the $j$'s Fourier frequency given by $f_s / N_s$ and $\hat{}$ indicates that each $\hat{P_i}(f_j)$ is an estimate of some true \gls{PSD} $P(f_j)$.
Finally, to obtain the standard \gls{WOSA} estimate, the average of the modified periodograms is computed at each frequency $f_j$.

In contrast to the \gls{WOSA} estimator, the \gls{WP} estimator computes the $q^\mathrm{th}$ sample quantile of the set $\lbrace \hat{P}_i(f_j)\rbrace$ for each $f_j$ (which is equivalent to the $p = q \cdot 100$ percentile).
To do so, first the order statistic $\lbrace \hat{P}_{(1)}, \, \dots , \hat{P}_{(K)} \rbrace$ is determined at each frequency bin. (We have dropped the dependence on $f_j$ for the sake of brevity.)
Afterwards, the $q^\mathrm{th}$ sample quantile can be computed according to \cite{par79} as

\begin{multline}\label{eq:sample_quantile}
    \hat{Q}(q) = K \left( \frac{i}{K} - q \right) \hat{P}_{(i-1)} + K \left(q -  \frac{i-1}{K} \right) \hat{P}_{(i)} \\ \mathrm{for} \quad \frac{i-1}{K} \leq q \leq \frac{i}{K} \quad \mathrm{and} \quad i = 1, \, \dots , K.
\end{multline}

\noindent That is, if the desired quantile falls between two samples $\hat{P}_{(i-1)}$ and $\hat{P}_{(i)}$, the sample quantile is estimated via linear interpolation.
As we will show in Section~\ref{sec:statistical_properties}, the sample quantile is, in general, biased compared to the true \gls{PSD}, whereby the bias $b$ depends on $q$ and $K$.
Hence, the final \gls{WP} estimator can be defined as

\begin{equation}\label{eq:def_welch_percentile}
    \hat{P}^{(\mathrm{WP})}_q = \frac{\hat{Q}(q)}{b(q,K)}.
\end{equation}



\section{Statistical Properties of the WP Estimator}
\label{sec:statistical_properties}

\subsection{Distribution}

The statistical properties of the \gls{WP} estimator can be derived from the order statistics $\lbrace \hat{P}_{(1)}, \, \dots , \hat{P}_{(K)} \rbrace$ of the modified periodograms.
Here, it is assumed that the $\Hat{P}_i$'s are independent and identically distributed\footnote{The independence condition holds if adjacent data segments do not overlap, or a moderate overlap along with a proper data taper is used.}.
It is well known (for example, see \cite[p.~224-225]{per20}) that for a proper window and large enough $N_s$ the distribution of $\Hat{P}_i$ is given by

\begin{equation}\label{eq:distribution_modified_periodogram}
    \hat{P}_i \overset{\mathrm{d}}{=} \frac{P}{2} \chi^2_2 \quad \mathrm{for} \quad 0 < f_j < \frac{f_s}{2}
\end{equation}

\noindent where $\chi^2_2$ is the chi-square distribution with two degrees of freedom and \gls{PDF}

\begin{equation}\label{eq:pdf_chi2}
    f(u) = \begin{cases}
               \frac{1}{2} \mathrm{e}^{-u/2}, & u \geq 0\\
               0, & u < 0
            \end{cases}.
\end{equation}

According to \cite{dav81}, the \gls{PDF} $f_{(i)}(x)$ of the $i^\mathrm{th}$ order statistic $\hat{P}_{(i)}$ is given by
\begin{equation}\label{eq:pdf_order_statistics}
    f_{(i)}(x) = \frac{1}{\mathrm{B}(i, K-i+1)} F^{i-1}(x) \left( 1 - F(x) \right)^{K-i} f(x),
\end{equation}

\noindent where $F(x)$ is the cumulative distribution function of $\hat{P}_i$ and can be obtained by integrating Equation~\eqref{eq:pdf_chi2} from \num{0} to $x$.
$\mathrm{B}(\alpha, \beta)$ is the beta function defined by

\begin{equation}\label{eq:beta_function}
    \mathrm{B}(\alpha, \beta) = \int_0^1 t^{\alpha-1} \left( 1-t \right)^{\beta-1} \mathrm{d}t.
\end{equation}


\noindent Equation~\eqref{eq:pdf_order_statistics} can now be used to derive expressions for bias and variance of the \gls{WP} estimator.

\subsection{Bias}\label{ssec:bias}

As shown in \cite{all12} for $q=0.5$ (i.e., the sample median) and for odd $K$ the bias can be expressed as
\begin{equation}\label{eq:bias_ahs}
    b = \sum_{k=1}^K \frac{\left( -1 \right)^{k+1}}{k}.
\end{equation}

\noindent Following their procedure, similar expressions for an arbitrary quantile and sample size can be derived.
To do so, we first substitute $\alpha = K-i$, $\beta = i-1$, and $t = 1 - F(x)$ and then compute the expected value of $\hat{Q}(q)$ using Equation~\eqref{eq:pdf_order_statistics}.
We also make the assumption that $\hat{Q}(q) \approx \hat{P}_{(i)}$ for some $i = 1, \, \dots, K$.
While this, in general, does not reflect the \gls{WP} estimator defined in Equation~\eqref{eq:sample_quantile} and~\eqref{eq:def_welch_percentile}, we have found that this approximation provides good results for the estimator's statistical properties.
The resulting $\mathbb{E}\lbrace \hat{Q}(q) \rbrace$ is given in Equation~\eqref{eq:quantile_expectation_integral}.



\begin{equation}\label{eq:quantile_expectation_integral}
    \mathbb{E}\lbrace \hat{Q}(q) \rbrace \approx -\frac{P}{\mathrm{B}(\alpha + 1, \beta + 1)} \int_0^1 t^{\alpha} (1-t)^{\beta} \ln(t) \mathrm{d}t.
\end{equation}

\noindent Here, we have used the fact that $\mathrm{d}t = -f(x) \mathrm{d}x$ and $x = - P \ln \left(1 - F(x) \right) = -P \ln(t)$ for the chi-square distribution with \num{2} degrees of freedom.
By noting that

\begin{equation}\label{eq:derivative_beta_fct}
    \frac{\partial t^\alpha (1-t)^\beta}{\partial \alpha} = t^\alpha (1-t)^\beta \ln(t),
\end{equation}

\noindent Equation~\ref{eq:quantile_expectation_integral} can be written as


\begin{equation}\label{eq:quantile_expectation_beta}
    \mathbb{E}\lbrace \hat{Q}(q) \rbrace = -\frac{P}{\mathrm{B}(\alpha + 1, \beta + 1)} \frac{\partial}{\partial \alpha} \mathrm{B}(\alpha + 1, \beta + 1).
\end{equation}



\noindent Using the digamma function $\psi$ to express the partial derivative of the beta function\footnote{$\frac{\partial \mathrm{B}(\alpha, \beta)}{\partial \alpha} = \mathrm{B}(\alpha, \beta) \left[ \psi(\alpha) - \psi(\alpha + \beta) \right]$}, Equation~\eqref{eq:quantile_expectation_beta} can be simplified to

\begin{equation}\label{eq:quantile_expectation_digamma}
    \mathbb{E}\lbrace \hat{Q}(q) \rbrace = P \left[ \psi(\alpha + \beta + 2) - \psi(\alpha + 1) \right].
\end{equation}

\noindent This shows that the bias between the \gls{WP} estimator and the true \gls{PSD} $P$ is given by

\begin{equation}\label{eq:bias_alpha_beta}
    b = \psi(\alpha + \beta + 2) - \psi(\alpha+1).
\end{equation}

\noindent Using the fact that $\psi$ can be expressed as

\begin{equation}\label{eq:digamma_harmonic_num}
    \psi(n) = - \gamma + \sum_{k=1}^{n-1} \frac{1}{k}, \quad \mathrm{for} \quad n \geq 2
\end{equation}

\noindent where $\gamma$ is the Euler-Mascheroni constant \cite{abr74}, the bias takes the form of a truncated harmonic series:

\begin{equation}\label{eq:bias_ths}
    b = \sum_{k = \alpha + 1}^{\alpha + \beta + 1} \frac{1}{k}, \quad \mathrm{for} \quad \alpha, \beta \in \mathbb{N}.
\end{equation}

To express the bias by means of $K$ and $q$, it is helpful to interpret $\alpha$ and $\beta$ as the number of samples with values greater and smaller than the desired percentile $\hat{P}_{(i)}$.
Therfore, we have to distinguish between two cases: (1) the sample percentile $\hat{Q}(q)$ matches exactly with one of the periodograms $\hat{P}_{(i)}$ (e.g., if $q=0.5$ and $K$ is odd), or (2) the sample percentile falls in between two periodograms $\hat{P}_{(i-1)}$ and $\hat{P}_{(i)}$ (e.g., if $q=0.5$ and $K$ is even).
In the former case, $\alpha$ and $\beta$ can be expressed by $\alpha = (K-1)(1-q)$ and $\beta = (K-1)q$, respectively.
For the latter case, $\alpha = K(1-q)$ and $\beta = Kq$ are natural choices.
Using this parameterization, Equation~\eqref{eq:bias_ths} can be rewritten as

\begin{equation}\label{eq:bias_ths_nq}
    b = \begin{cases}
               \sum\limits_{k = (K-1)(1-q)+1}^{K} \frac{1}{k}, & \hat{Q}(q) =  \hat{P}_{(i)} \vspace{1em}\\
               
               \sum\limits_{k = K(1-q)+1}^{K+1} \frac{1}{k}, & \hat{P}_{(i-1)} < \hat{Q}(q) <  \hat{P}_{(i)}
            \end{cases}
\end{equation}

\noindent In the limit, that is, for $K \longrightarrow \infty$ both cases converge to $-\ln(1-q)$.
Furthermore, the products $(K-1)(1-q)$ and $K(1-q)$ have to be integers, or, otherwise, rounded to the next nearest integer to compute the bias.
If the constellation of $K$ and $q$ does not result in an integer value, the polynomial approximation for the digamma function \cite{abr74}

\begin{equation}\label{eq:approx_digamma}
    \psi(n) \approx \ln(n) - \frac{1}{2n} - \frac{1}{12n^2} + \frac{1}{120n^4} - \frac{1}{252n^6}
\end{equation}

\noindent can be used to avoid rounding.
In this case, the bias should be computed by

\begin{equation}\label{eq:bias_dga}
    b = \psi(K+2) - \psi(K(1-q) + 1).
\end{equation}

\subsection{Variance}\label{ssec:variance}

In analogy to Equation~\eqref{eq:quantile_expectation_integral}, the second order moment of the sample quantile is given by



\begin{equation}\label{eq:quantile_var_integral}
    \mathbb{E}\lbrace \hat{Q}^2_{(q)} \rbrace = \frac{P^2}{\mathrm{B}(\alpha + 1, \beta + 1)} \int_0^1 t^{\alpha} (1-t)^{\beta} \left[\ln(t)\right]^2 \mathrm{d}t.
\end{equation}

\noindent By using the relation

\begin{equation}\label{eq:second_derivative_beta_fct}
    \frac{\partial^2 t^\alpha (1-t)^\beta}{\partial \alpha^2} = t^\alpha (1-t)^\beta \left[\ln(t)\right]^2
\end{equation}

\noindent and taking the second derivative of the beta function with respect to $\alpha$, the second order moment yields



\begin{multline}\label{eq:quantile_var_digamma_trigamma}
    \mathbb{E}\lbrace \hat{Q}^2_{(q)} \rbrace = P^2 \left( \left[ \psi(\alpha+1) - \psi(\alpha + \beta + 2) \right]^2 \right. \\ \left.  + \left[ \psi_1(\alpha + 1) - \psi_1(\alpha + \beta + 2) \right]\vphantom{\left[\right]^2} \right).
\end{multline}

\noindent Therein, $\psi_1(n)$ is the derivative of the digamma function -- also referred to as trigamma function -- and can be approximated by means of Equation~\eqref{eq:approx_digamma} as

\begin{equation}\label{eq:trigamma_fct}
    \psi_1(n) = \frac{\mathrm{d}\psi(n)}{\mathrm{d}n} \approx \frac{1}{n} + \frac{1}{2n^2} + \frac{1}{6n^3} - \frac{1}{30n^5} + \frac{1}{42n^7}.
\end{equation}

\noindent From the first and second order moments of the sample quantile, the variance of the \gls{WP} estimator can be determined:

\begin{equation} \label{eq:quantile_var_trigamma}
    \mathrm{var}\lbrace \hat{P}_q^{(\mathrm{WP})} \rbrace = \frac{P^2}{b^2} \left[ \psi_1(\alpha + 1) - \psi_1(\alpha + \beta + 2) \right].
\end{equation}

\noindent For the general case, i.e., if $\hat{P}_{(i)} \neq \hat{Q}(p)$, the \gls{WP} estimator's variance can be computed by

\begin{equation} \label{eq:quantile_var_trigamma_q}
    \mathrm{var}\lbrace \hat{P}_q^{(\mathrm{WP})} \rbrace = \frac{P^2}{b^2} \left[ \psi_1(K(1-q) + 1) - \psi_1(K + 2) \right].
\end{equation}

\subsection{Limiting Distribution}\label{ssec:limiting_distribution}

For $K \rightarrow \infty $, the order statistic of the modified periodograms is normally distributed around $-P \ln(1-q)$ with variance 

\begin{equation}\label{eq:lim_variance_quantile}
    \mathrm{var}\lbrace \hat{Q}(q) \rbrace = \left(\frac{P}{2}\right)^2 \cdot \frac{q(1-q)}{K f^2(-2 \ln(1-q))},
\end{equation}

\noindent where $f$ is the \gls{PDF} given in Equation~\eqref{eq:pdf_chi2} \cite{ken94}.
Simplifying this expression and taking the bias correction into account, the limiting variance of the \gls{WP} estimator can be computed by

\begin{equation}\label{eq:lim_variance_wp}
    \mathrm{var}\lbrace P_q^{(\mathrm{WP})} \rbrace = \left(\frac{P}{b}\right)^2 \cdot \frac{q}{K (1-q)},
\end{equation}

\subsection{Equivalent Degree of Freedom}\label{ssec:edof}

So far, we have assumed that adjacent periodograms are approximately independent.
Now, we want to relax this condition by introducing the concept of \gls{EDOF} to get the number of independent random variables of the quantile estimation.
According to \cite[p.~429]{per20}, the \gls{EDOF} $\nu$ for the \gls{WOSA} estimator is

\begin{equation}\label{eq:edof}
    \nu = \frac{2K}{1 + 2 \sum\limits_{m=1}^{K-1} \left( 1 - \frac{m}{K} \right) \left\vert \sum\limits_{t=0}^{N_\mathrm{s}-1} h_t h_{t+mN_\mathrm{o}} \right\vert},
\end{equation}

\noindent where $h_t$ is the data taper and $N_\mathrm{o}$ is the number of overlapping samples.
Since the same periodograms are used for the \gls{WOSA} and \gls{WP} estimator, Equation~\eqref{eq:edof} also holds for the latter one.
That is, the \gls{WP} estimator uses $\nu/2$ equivalent and independent periodograms to estimate the true \gls{PSD}.
Bias, variance, and limiting variance can now be computed for arbitrary overlaps and data taper when $K$ is replaced by $\nu/2$.
(Note that in Equation~\eqref{eq:bias_ths_nq}, $\nu/2$ and the product $\frac{\nu}{2}(1-q)$ would need to be rounded to the next nearest integers.)

\section{Simulations}\label{sec:simulations}

Here, we will compare the previously derived expressions for bias and variance with results from a simulated white Gaussian noise sequence.
All data segments have a length of $N_\mathrm{s} = 1024$ and a Hann data taper with \SI{50}{\percent} overlap is used.
Subsequently, the \gls{WP} estimate according to Equation~\eqref{eq:sample_quantile} and~\eqref{eq:def_welch_percentile} is computed for various $q$ and $K \geq 3$, and sampling bias and variance are calculated.
To reduce the variability in the estimate, \num{51100} independent trials are averaged for each $K$ and $q$.
Furthermore, the \gls{EDOF} instead of $K$ is used in all formulas. 
It is noted that the goodness of fit between simulations and theoretical results is independent of the data taper if the \gls{EDOF} is used.
This has been tested using the Slepian, Parzen, and triangular window.

\begin{figure}[htb]

\centerline{\includegraphics[width=8.5cm]{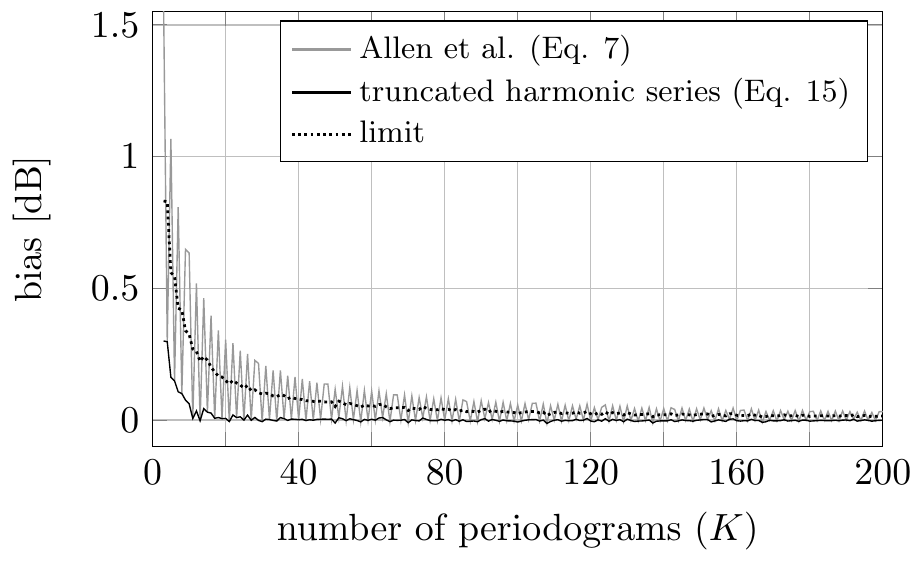}}
\caption{Bias of the Welch $50^\mathrm{th}$ percentile estimator after correcting the quantile bias according to Equation~\eqref{eq:bias_ahs} (Allen et al.), \eqref{eq:bias_ths_nq} (truncated harmonic series), and $b = - \ln(0.5)$ (limit).}
\label{fig:bias_median}
\end{figure}

Figure~\ref{fig:bias_median} shows the the bias of the Welch $50^\mathrm{th}$ percentile estimator after applying the bias correction using Equation~\eqref{eq:bias_ahs}, \eqref{eq:bias_ths_nq}, and the limit $b=-\ln(0.5)$.
If $\nu/2$ rounded to the next nearest integer is odd, Equation~\eqref{eq:bias_ahs} and~\eqref{eq:bias_ths_nq} give identical results with bias values smaller than \SI{0.1}{\decibel} for $K \geq 7$.
However, if $\nu/2$ is even, only Equation~\eqref{eq:bias_ths_nq} is capable of accurately compensating the bias.
Note that the rounded $\nu/2$ is in general not equal to $K$ for the given window and overlap. 
When using the limit of the bias ($b=-\ln(0.5)$) equally good results for even and odd $K$ are obtained, but the performance is worse compared to Equation~\eqref{eq:bias_ths_nq}.
In general the truncated harmonic series is favorable as it gives the lowest bias over all $K$.
However, for sufficiently large $K$, accurate results can be achieved for all three bias correction expressions.

\begin{figure}[htb]

\centerline{\includegraphics[width=8.5cm]{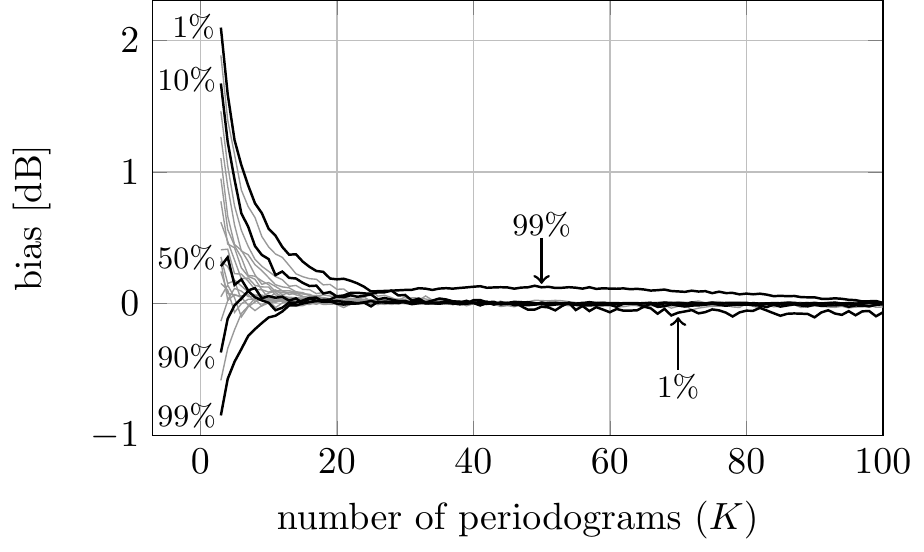}}
\caption{Bias of the \gls{WP} estimator for percentiles between \SI{1}{\percent} and \SI{99}{\percent} in \SI{5}{\percent} increments after correcting the quantile bias using the digamma approximation (Equation~\eqref{eq:approx_digamma} and~\eqref{eq:bias_dga}).}
\label{fig:bias_percentile}
\end{figure}

The bias of the \gls{WP} estimator for different percentiles is shown in Figure~\ref{fig:bias_percentile}.
Therein, the digamma approximation (Equation~\eqref{eq:approx_digamma} and~\eqref{eq:bias_dga}) is used to compensate for the quantile bias. 
One can observe that, for small values of $K$, more extreme percentiles tend to over or underestimate the true \gls{PSD}, whereas percentiles around $\SI{63}{\percent}$ (unbiased estimator) exhibit only a small or no bias.
Only for the $1^\mathrm{st}$ and $99^\mathrm{th}$ percentile a bias greater than \SI{0.1}{\decibel} can still be observed for some $K \geq 30$. (This bias will also vanish as $K$ further increases.)

\begin{figure}[htb]

\centerline{\includegraphics[width=8.5cm]{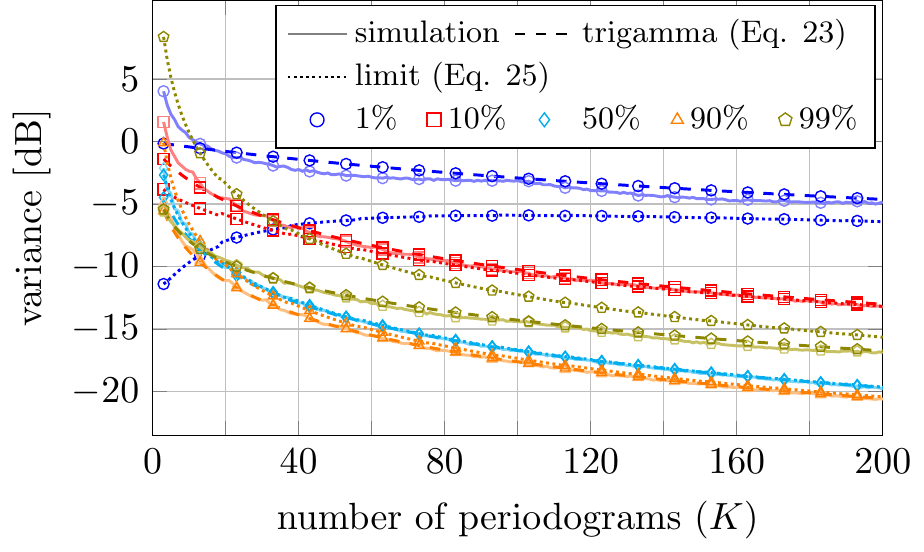}}
\caption{Simulated and theoretical variance according to Equation~\eqref{eq:quantile_var_trigamma_q} (trigamma) and Equation~\eqref{eq:lim_variance_wp} (limit) of the \gls{WP} estimator.}
\label{fig:var_percentile}
\end{figure}

Finally, variance and limiting variance according to Equation~\eqref{eq:quantile_var_trigamma_q} and~\eqref{eq:lim_variance_wp} are compared to the sampling variance of the \gls{WP} estimator in Figure~\ref{fig:var_percentile}.
The bias is corrected using the digamma approximation.
The results show that Equation~\eqref{eq:quantile_var_trigamma_q} deviates by less than \SI{0.5}{\decibel} from the simulations for $K \geq 16$ and percentiles between \SI{10}{\percent} and \SI{90}{\percent}.
The limiting variance (Equation~\eqref{eq:lim_variance_wp}), on the other hand, requires values $K \geq 79$ to provide the same accuracy.
For the $1^\mathrm{st}$ and $99^\mathrm{th}$ percentile, a greater deviation between theoretical expressions and simulations can be observed.
In these cases, larger values of $K$ would be necessary to achieve a better fit.
Figure~\ref{fig:var_percentile} also shows that the variance of the $50^\mathrm{th}$ percentile estimator (median) is larger compared to the variance of the $90^\mathrm{th}$ percentile estimator.
Indeed one can show that, in the limit, the $80^\mathrm{th}$ percentile estimator has the lowest variance -- by a factor of approximately \SI{1.3}{\decibel} compared to the median -- among all \gls{WP} estimators.

\section{Conclusion}\label{sec:conclusion}

The \gls{WP} estimator is a robust approach for computing \gls{PSD} estimates.
Equations for the bias of the underlying quantile estimate have been derived.
We have shown that our bias correction approach outperforms the existing method for the Welch $50^\mathrm{th}$ percentile estimator and also performs excellent for other percentiles.
Furthermore, simple expressions for the estimator's variance have been derived and comparisons with simulated data have shown great agreement for most percentiles and a wide range of sample sizes.


\bibliographystyle{IEEEbib}
\bibliography{refs}

\begin{thebibliography}{10}

\bibitem{wel67}
P.~{Welch},
\newblock ``The use of fast fourier transform for the estimation of power
  spectra: A method based on time averaging over short, modified
  periodograms,''
\newblock {\em IEEE Transactions on Audio and Electroacoustics}, vol. 15, no.
  2, pp. 70--73, 1967.

\bibitem{mar19}
R.~A. Maronna, R.~Martin, V.~J. Yohai, and M.~Salibián-Barrera,
\newblock {\em Robust Statistics, 2nd Edition},
\newblock Wiley, 2nd edition, 2019.

\bibitem{all12}
B.~Allen, W.~G. Anderson, P.~R. Brady, D.~A. Brown, and J.~D.~E. Creighton,
\newblock ``Findchirp: An algorithm for detection of gravitational waves from
  inspiraling compact binaries,''
\newblock {\em Physical Review D}, vol. 85, no. 12, Jun 2012.

\bibitem{par09}
S.~E. Parks, I.~Urazghildiiev, and C.~W. Clark,
\newblock ``Variability in ambient noise levels and call parameters of north
  atlantic right whales in three habitat areas,''
\newblock {\em The Journal of the Acoustical Society of America}, vol. 125, no.
  2, pp. 1230--1239, 2009.

\bibitem{mer12}
N.~D. Merchant, P.~Blondel, D.~T. Dakin, and J.~Dorocicz,
\newblock ``Averaging underwater noise levels for environmental assessment of
  shipping,''
\newblock {\em The Journal of the Acoustical Society of America}, vol. 132, no.
  4, pp. EL343--EL349, 2012.

\bibitem{kee17}
J.~T. {Kees}, J.~M. {Ernst}, W.~C. {Headley}, and A.~A. {Louis Beex},
\newblock ``Robust blind spectral estimation in the presence of non-gaussian
  noise,''
\newblock in {\em MILCOM 2017 - 2017 IEEE Military Communications Conference
  (MILCOM)}, Baltimore, MD, USA, 2017, pp. 629--634.

\bibitem{nut71}
A.~H. {Nuttail},
\newblock ``Spectral estimation by means of overlapped fast fourier transform
  processing of windowed data,''
\newblock Tech. {R}ep. 4169, NUSC, New London, CT, USA, 1971.

\bibitem{per20}
D.~B. Percival and A.~T. Walden,
\newblock {\em Spectral Analysis for Univariate Time Series},
\newblock Cambridge Series in Statistical and Probabilistic Mathematics.
  Cambridge University Press, Cambridge, UK, 2020.

\bibitem{par79}
E.~Parzen,
\newblock ``Nonparametric statistical data modeling,''
\newblock {\em Journal of the American Statistical Association}, vol. 74, no.
  365, pp. 105--121, 1979.

\bibitem{dav81}
H.~A. David and H.~N. Nagaraja,
\newblock ``Distribution of a single order statistic,''
\newblock in {\em Order Statistics}, pp. 9--11. Wiley, 3rd edition, 2004.

\bibitem{abr74}
M.~Abramowitz and I.~Stegun,
\newblock ``Psi (digamma) function,''
\newblock in {\em Handbook of mathematical functions with formulas, graphs, and
  mathematical tables}, Applied mathematics series (Washington, D.C.); 55, p.
  258. U.S. Dept. of Commerce : U.S. G.P.O., Washington, D.C., USA, 10th
  print., with corrections. edition, 1972.

\bibitem{ken94}
M.~G. Kendall, A.~Stuart, J.~K. Ord, S.~F. Arnold, and A.~O'Hagan,
\newblock ``Standard errors of quantiles,''
\newblock in {\em Kendall's advanced theory of statistics}, Kendall's library
  of statistics, pp. 356--358. Edward Arnold, London, UK, 6th edition, 1994.

\end{thebibliography}

\end{document}